\documentclass{svproc}
\usepackage{url}

\usepackage{graphicx}
\usepackage{subfigure}
\DeclareGraphicsExtensions{.png,.eps,.ps,.pdf}

\begin{document}
\mainmatter              
\title{Drowsiness detection in drivers with a smartwatch}
\titlerunning{Drowsiness detection} 
\author{Sonia Díaz-Santos\inst{1}, Pino Caballero-Gil\inst{2}}
\authorrunning{Sonia Díaz-Santos, Pino Caballero-Gil} 
\tocauthor{Sonia Díaz-Santos, Pino Caballero-Gil}
\institute{University of La Laguna, St. San Francisco de Paula, 19, 38200, San Cristóbal de La Laguna, Tenerife, Spain\\
\email{sdiazsan@ull.edu.es, pcaballe@ull.edu.es}}
\maketitle             

\begin{abstract}

The main objective of this work is to detect early if a driver shows symptoms of sleepiness that indicate that he/she is falling asleep and, in that case, generate an alert to wake  him/her up. To solve this problem, an application has been designed that collects various parameters, through a smartwatch while driving. First, the application detects the driving action. Then, it collects information about the most significant physiological variables of a person while driving. On the other hand, given the high level of sensitivity of the data managed in the designed application, in this work special attention has been paid to the security of the implementation. The proposed solution improves road safety, reducing the number of accidents caused by drowsiness while driving.

\end{abstract}
\section{Introduction}

Traffic accidents often endanger the lives not only of the driver but also of other people on the road. It is therefore necessary to do everything possible to reduce their number. Among the options to achieve this, this work has opted for the development of innovative technologies to address the problem. Specifically, the present study has focused on identifying physiological variables that characterize sleepiness or fatigue while driving, with the aim of using them to reduce the number of accidents caused by that reason, since, in general, drowsiness is involved, directly or indirectly, in 15-30\% of traffic accidents.

In this work, a study of the most relevant physiological variables that allow us to conclude whether a person is falling asleep has been carried out. As discussed in this paper, according to various publications, these variables are: heart rate, electrocardiogram, respiratory function and stress. 
Specifically, by means of the smartwatch used in this work, it has been possible to collect the data of some of these physiological variables of a person, to locate the signs of drowsiness and check, in real time, if the person is falling asleep \cite{nav:agui}.

On the other hand, to detect the action of driving, the smartwatch has a set of sensors such as accelerometer, gyroscope, pedometer and Global Positioning System (GPS).

Thus, this work starts from the intersection between the list of physiological variables descriptive of sleep and that of the sensors available in the used smartwatch. In particular, some parameters such as Heart Rate Variability (HRV) have been used. Then, these collected data have been analyzed to detect drowsiness in drivers. For this purpose, different sensors such as PhotoPlethysmoGraphy (PPG) and ElectroCardioGram (ECG) sensors have been used. The PPG sensor uses a light-based technology to detect the rate of blood flow controlled by the pumping action of the heart. In addition, other sensors such as the accelerometer, gyroscope, pedometer and GPS, integrated into the watch, have been used to monitor the user's physical activity \cite{bece:jhaq:vel:tona:rami}.
                 
Other factors have also been taken into account, such as the time of day, allowing for the circadian rhythm according to which physical, mental and behavioral changes occur in 24-hour cycles. In fact, it is known that fatigue-related accidents are strongly dependent on the time of day, as well as on the type of road, especially on monotonous roads.
In addition, other factors to consider are a person's age, gender and regular physical exercise, to determine the normal parameters for each individual. For example, heart rate in athletes is usually lower than in sedentary people, and heart rate is usually lower in women than in men. Keeping all these indicators in mind, a better analysis of the drowsiness data has been possible \cite{vic:lag:bar}.\par
This document is structured as follows. Section II provides data from the used smartwatch and platform. Section III contains a discussion of physiological variables and sensors of interest for this work. Section IV provides a description of the design and functionalities of the proposed application. Section V describes some features of the secure implementation of the application. Finally, section VI closes the paper with some conclusions and future work.
\section{Smartwatch and platforms used}
This section provides some details of the  smartwatch chosen to measure the driver data, as well as of the platform used to develop the applications \cite{lee:lee:mill:chung}.
\subsection{Hardware}
For the development of this work, the smartwatch {\it Samsung Galaxy Watch 4 LTE} has been used. This watch has the  {\it Wear OS Powered by Samsung}, which allows health monitoring 24 hours a day. It has a sensor {\it BioActive} that measures ECG and blood pressure in real time. To measure blood pressure, it uses an optical PPG heart rate sensor, and to measure ECG, it uses an electrical heart sensor. It also allows measuring body composition through the Bioelectrical Impedance Analysis (BIA) sensor. This device allows blood oxygen and stress levels to be measured to obtain a complete sleep analysis. It has two motion sensors, the accelerometer and the gyroscope, which allow to know the location with the GPS sensor, and calculates the steps with the pedometer sensor. In terms of connections, it has Bluetooth 5.0 and Wi-Fi connection \cite{li::lee:chung}.
\subsection{Software}
This section defines the software used by the devices, including the communication architecture between the systems, as well as the technologies and development environments. It also mentions the Health Platform and the Wear OS community.
\subsubsection{Architecture communication scheme} 
\begin{figure}[htpb]
\centerline{
    \includegraphics[width=7.3cm]{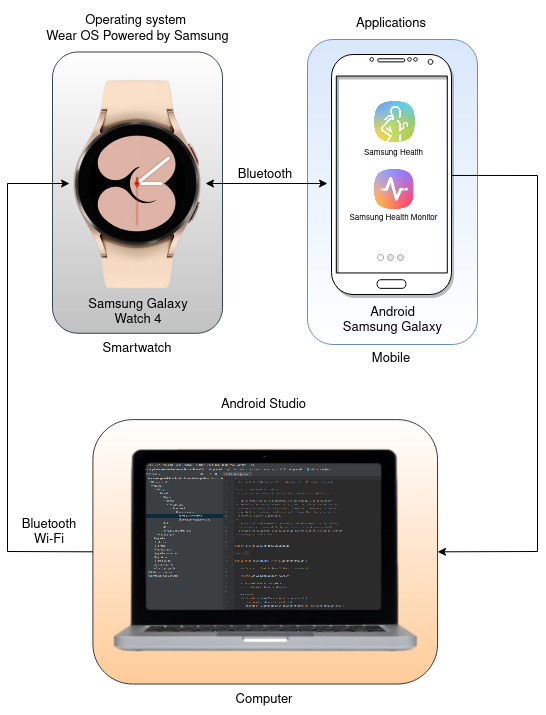}
}
    \caption{Communication scheme}
    \label{fig:Architecture_communication_scheme}
\end{figure}

Bluetooth or Wi-Fi connections are required for communication between the different devices, as shown in Figure \ref{fig:Architecture_communication_scheme}. The smartwatch with the {\it Wear OS Powered by Samsung}  has a Bluetooth connection to connect to the cell phone. The cell phone has the application {\it Samsung Health}, which collects the necessary data of physiological variables, as well as the application {\it Samsung Health Monitor} to obtain the data of the blood pressure and electrocardiogram of the person using the watch. The application is created on the computer and installed on the smartwatch through the development environment {\it Android Studio}. The application is installed on the watch through the aforementioned connections.
In addition, with {\it One UI Watch}, compatible apps are automatically installed on the watch when they are downloaded to the cell phone.
\subsubsection{Technologies} 
The used technologies  include the {\it Android Studio} tool to develop the application. The programming languages that can be used to carry it out are {\it Kotlin} and {\it Java}. It is necessary to have the applications in the cell phone to have access to the driver's data and perform its analysis in order to detect if the driver is driving and is having symptoms of drowsiness.
\subsubsection{Health Platform} 
\begin{figure}[ht]
\centerline{
    \includegraphics[width=5cm]{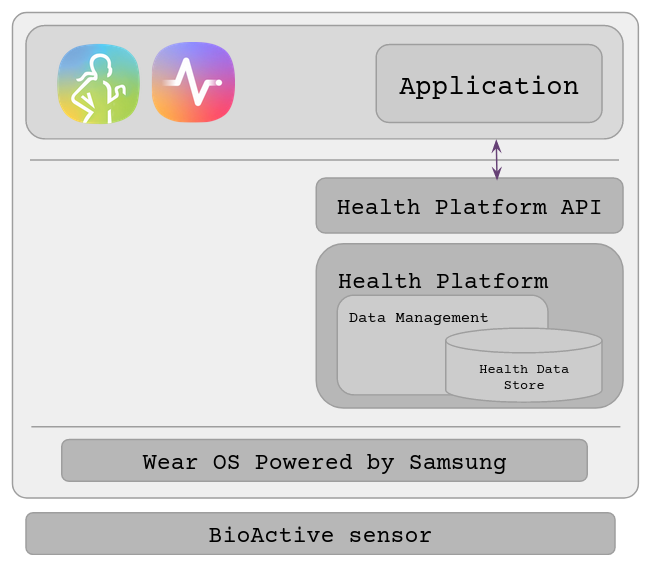}
}
    \caption{Application scheme}
    \label{fig:App_scheme}
\end{figure}

The {\it Samsung Health} Software Development Kit (SDK) for Android enables the sharing of health data between {\it Samsung Health}, which runs on Android phones, and partner applications, as shown in Figure \ref{fig:App_scheme}. It also allows partner applications to use the tracking feature of {\it Samsung Health} by creating applications with the SDK.
The SDK provides secure access to {\it Samsung Health} data with the applicable data types. However, data sharing is only enabled after explicit consent from the user. The user can select detailed data sharing settings, including which partner application will access the user's data and what type of data will be read or written.
\subsubsection{Wear OS Community} 
The {\it Wear OS Powered by Samsung} community has a dedicated platform for Android developers with a multitude of tutorials about any type of Android device. There are communities on YouTube: {\it Android Developers}, on Twitter: {\it AndroidDev} and on LinkedIn: {\it Android Developers}.
\subsection{Compatibility}
Compatibility between devices to have full functionality and access to all data is complex. The smartwatch {\it Samsung Galaxy Watch 4} is compatible with any Android device, but in order to obtain the data from the ECG and blood pressure sensors, the cell phone must be a {\it Samsung Galaxy} with an Android version higher than N, a RAM memory capacity of at least 1.5 Gb and have Google Mobile Services (GMS). This means that although the {\it Samsung Health} application is compatible with Android devices, including non-Samsung devices, if you want to get that physiological data and install the {\it Samsung Health Monitor} application, the restrictions change.
\section{Physiological variables and sensors}
This section details how the physiological variables most relevant to this work are influenced and how the sensors used work to collect the data \cite{ram::khan:awan:ism:ily:mah}.
\subsection{Sleep phases}
\begin{figure}[ht]
\centerline{
    \includegraphics[width=7cm]{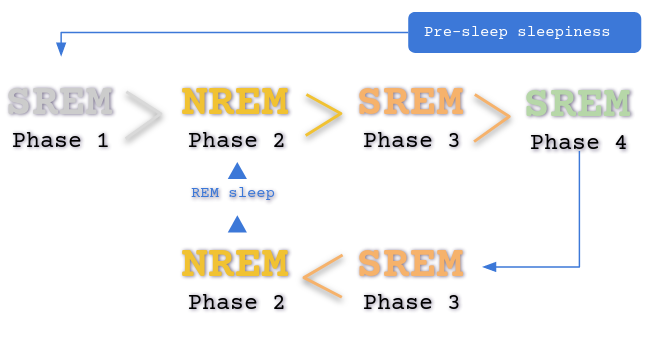}
}
    \caption{Diagram of the phases of sleep}
    \label{fig:Sleep_phases}
\end{figure}

Figure \ref{fig:Sleep_phases} shows the schematic of sleep phases, in which initial drowsiness is the most relevant for this work. 
Phase 1 is the lightest degree of sleep and lasts a few minutes. In this phase, physiological activity decreases with a gradual drop in vital signs and metabolism. In addition, in this phase it is easy to be awakened by sensory stimuli.

During sleep, not only the best-known changes occur, such as ElectroEncephaloGram (EEG) alterations, Rapid Eye Movements (REM) or muscle tone alterations  assessed with ElectroMyoGraphy (EMG), but also important cardiovascular, respiratory, hormonal, renal, digestive and general changes of the whole organism. Blood pressure (5-16\%), pulse and respiratory function are decreased.
\subsection{Sleep and driving detection}

To detect sleep, data on the following physiological variables are collected: heart rate, stress, blood pressure and blood oxygen \cite{dew::zul}.

Resting heart rate is the number of times  the heart beats per minute when  at rest. A low resting heart rate is usually synonymous with good cardiovascular health. Aerobic exercise can help lower the resting heart rate over time. Temperature, body position, recent activity or emotional state are some of the factors that can affect heart rate. In the {\it Samsung Health} application, resting heart rate data is based on estimates from U.S. residents. 

Stress is measured with certain biomarkers. The greater the number of measurements taken, the greater the accuracy of the stress data collected. Tobacco, alcohol, caffeine and medications can affect stress level measurements. When using the stress level function, the watches use heart rate data, such as beats per minute, to determine the interval between each heartbeat. Lower variability between beats equates to higher stress levels, while increased variability indicates less stress.

Blood pressure is measured by an optical heart rate sensor known as a PPG sensor. Monitoring your blood pressure is very important for your health. If your blood pressure is within the normal range, it is a good indication that you have a healthy heart. But having high blood pressure, also known as hypertension, can significantly increase the risk of brain, kidney and heart disease, including stroke and coronary heart disease when not managed properly. In the application they will be classified as: pulse, systolic blood pressure and diastolic blood pressure. The measurement range for blood pressure readings is: systolic: 70-180 and diastolic: 40-120.

The blood oxygen level is an indicator of how efficiently oxygen is transported through the body, which in turn indicates whether you are breathing efficiently. Blood oxygen level, also known as SpO2 (percutaneous oxygen saturation), is the measure of the percentage of hemoglobin that is oxygenated in the red blood cells. A healthy range is 95-100\% when at rest. Factors such as intense exercise, the amount of oxygen in the air, altitude and various health problems can give lower percentage readings \cite{tat::guan:cao:qu},  \cite{jo:kim:kim}.

All of this data is collected by the {\it Samsung BioActive} sensor, which is a 3-in-1 single chip sensor. The sensors are:  PPG, ECG and BIA.

The PPG sensor is capable of measuring heart rate, blood oxygen, stress level and respiratory rate. Photo means light, Plethysmograph means volume change, and Gram means graph. Therefore, PPG is a low intensity, high precision green infrared light sensor used to detect blood flow volume to understand the fluctuation in heart rate.

The ECG sensor is used for heart rate and rhythm detection. Since this sensor is bulky, it cannot be used to detect heart rate when the body is in motion. Thus, heart rate and Heart Rate Variability (HRV) can be measured accurately and continuously, even during extreme physical activity \cite{fuj}, \cite{vic:lag:bar:bai}, \cite{bue:for:kar:arne:anu:can}.

Functionally, when the heart beats, capillaries expand and contract according to changes in blood volume. PPG's optical sensor, which uses motion-tolerant technology, emits light signals that reflect off the skin to accurately and continuously measure weak blood flow signals. So when light travels through biological tissues, it is absorbed by bone, skin pigments, and venous and arterial blood. Since blood absorbs light more strongly than surrounding tissues, PPG sensors can detect changes in blood flow as changes in light intensity. The PPG voltage signal is proportional to the amount of blood flowing through the blood vessels. Even small changes in blood volume can be detected with this method, providing greater accuracy.

The electrocardiogram function works by recording the heart's electrical activity via a sensor on a compatible {\it Samsung Galaxy Watch}. The application measures heart rate and rhythm, which are classified as sinus rhythm or atrial fibrillation. A sinus rhythm is when the heart beats steadily. This occurs when the upper and lower chambers of the heart pump synchronously. Atrial fibrillation is when the heart beats in an irregular rhythm. This occurs when the upper chambers of the heart do not pump in sync with the lower chambers. If left untreated, it can lead to blood clots, strokes, heart failure and other health problems. If symptoms occur, they may include rapid heartbeat or palpitations, skipped beats, fatigue, shortness of breath, chest pressure or pain, fainting or dizziness \cite{awa:bad:dri}.

The BIA sensor performs real-time body composition analysis by placing two fingers on the two side buttons, which act as electrodes, to measure muscle mass, fat mass, body fat, Body Mass Index (BMI) and body water.

To detect the driving action, data is collected from the following sensors: accelerometer, gyroscope, pedometer and GPS.

\begin{figure}[ht]
\centerline{
    \includegraphics[width=6cm]{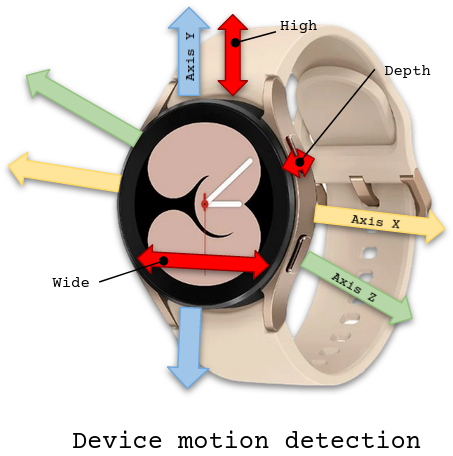}
}
    \caption{Accelerometer}
    \label{fig:Accelerometer}
\end{figure}

The accelerometer measures the force, direction and gravity of the acceleration and orientation of the device., as seen in Figure \ref{fig:Accelerometer}.

\begin{figure}[ht]
\centerline{
    \includegraphics[width=6cm]{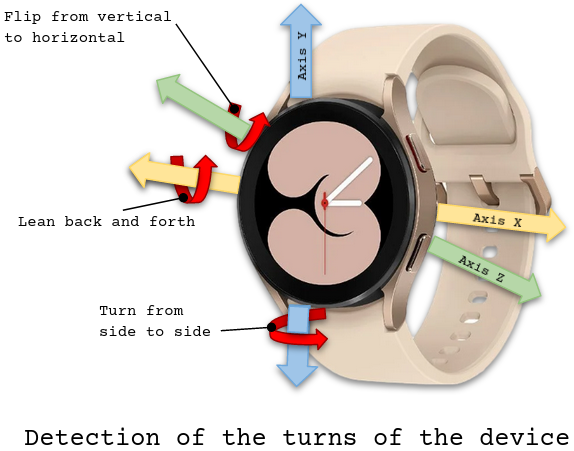}
}
    \caption{Gyroscope}
    \label{fig:Gyroscope}
\end{figure}

The gyroscope measures the angular velocity of the device and thus its exact position, as shown in Figure \ref{fig:Gyroscope}. By linking the data from the gyroscope with that from other sensors such as the accelerometer, the wristband maintains the correct orientation when you move, as well as being able to differentiate what type of movement you are making.

The pedometer counts the steps taken by the person wearing the watch. It also makes use of GPS to obtain much more accurate information. Since it is going to be able to measure the distance traveled, the time it has taken and the exact number of steps that have had to be taken to cover the distance in question. 

GPS is a global positioning system that allows you to determine the position of someone or something in precise latitude and longitude coordinates at any point on the planet in real time. The GPS receiver collects data from different satellites to calculate your position as a set of coordinates. This allows you to track your path and distance. In the specific case of sports, a GPS heart rate monitor will help you know your running speed, pace and distance traveled when you exercise.
\section{Application design and functionalities}
This section defines the objective, idea and design of the proposed application. In addition, the sensor variables of the platform {\it Android Developers} are described to understand how the physiological variables data are collected. Finally, the analysis of the data to determine whether a driver is falling asleep or not is discussed.
\subsection{Objective}
\begin{figure}[ht]
\centerline{
    \includegraphics[width=8cm]{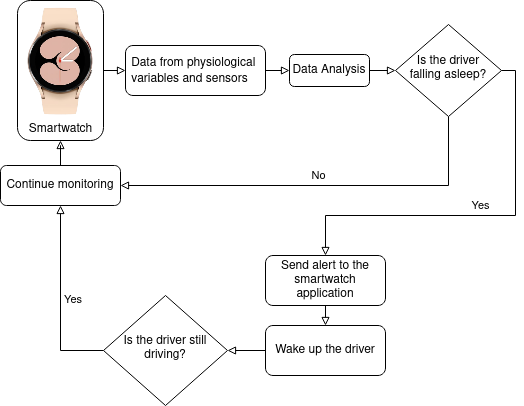}
}
    \caption{Application design}
    \label{fig:App_design}
\end{figure}

Figure \ref{fig:App_design} shows the use case diagram of the application, which describes the process of extracting data from the sensors and variables of the smartwatch. Then, the data is analyzed to determine if the driver shows symptoms of drowsiness. If the driver does not show symptoms of drowsiness and continues driving, monitoring is continued. If they do, a vibration alert is sent to the watch to wake up the driver. Finally, if the driver continues driving, the physiological variables continue to be monitored with the sensors.
\subsection{Sensors data}
The {\it Android Developers} platform has a {\it Sensor} public class whose purpose is to define the sensor variables in order to get the data from the smartwatch. There are several sensors that allow you to monitor the movement of a device: the vector sensors for rotation, gravity, linear acceleration, significant motion, step counter and step detector are either hardware or software based, and the accelerometer and gyroscope sensors are always hardware based. The most important sensor variables are divided into packages to store the classes. In the {\it android.hardware} package the most important variables are: {\it type accelerometer}, {\it type gyroscope}, {\it type heart rate}, {\it type heart beat} and {\it type step counter}. All of these variables have two data types {\it string} or {\it int}. The {\it android.location} package has the {\it Location} class to represent the geographic location in the variable {\it geographic location}. The {\it android.os} package contains the {\it Vibrator} class to produce the alarm on the clock. These are some of the examples of classes and variables used in this work.
\subsection{Data analysis}
Data can be analyzed with algorithms using two types of learning: supervised and unsupervised. Supervised learning  starts with a predefined set of  labeled data, so that the value of the  target attribute  is the attribute that you try to predict. target attribute for the data set is known at hand. On the other hand, unsupervised learning starts from previously unlabeled data.
This work initially assumes that there is no prior knowledge of a person's data, and therefore an unsupervised algorithm is used in this case. In addition, the case of measuring the data of interest in the resting state is considered in order to use a supervised algorithm.
\section{Application security}
By protecting the security of the application, users' confidence in it is enhanced. Therefore, in its development, we have tried to follow good security best practices in software implementation. 

Firstly, we have tried to ensure that the communication is secure, protecting the data exchanged by the application with other applications and web sites, improving the stability of the communication. This is achieved by using implicit {\it intents} and non-exported content providers (displaying an application selector, applying signature-based permissions, and disabling access to the application's content providers), requesting credentials before displaying sensitive information (by PIN,  password, pattern, or biometric credential, such as facial recognition or fingerprint), applying network security measures (using TLS traffic, adding a network security configuration, and creating a proprietary trust manager), and using {\it WebView} objects carefully (via HTML message channels). 
Secondly, care has been taken to ensure that permission requests are appropriate, using {\it intents} to defer permissions, and sharing data securely between applications. 
Third, attention has been paid to data storage, saving private data in internal storage, storing only non-sensitive data in cache files, and using {\it SharedPreferences} in private mode.
Fourth, the services and dependencies have been monitored for updates by checking the Google Play Services security provider and updating all application dependencies.

Also, the use of {\it SafetyNet} provides a set of services and APIs that help protect the application against security threats, which includes device tampering, incorrect URLs, potentially harmful applications, and fake users.
It features the {\it Android Keystore} system that protects the keying material from unauthorized use. This prevents the extraction of key material from application processes and the Android device as a whole in order to reduce unauthorized use of keys outside the device. In addition, it allows applications to specify authorized uses of their keys and enforces these restrictions outside of application processes to reduce unauthorized use of key material on the device.
\subsection{Key management}
Two concepts are used for key management. On the one hand, a set of keys is used to encrypt files or shared preference data, which are stored in {\it SharedPreferences}. On the other hand, a master key is used, which encrypts all key sets, and is stored using the Android keystore system.
The security library also includes two classes to provide more secure data at rest. First, the {\it EncryptedFile} class is used to provide secure read and write operations from file streams, using Authenticated Encryption with Associated Data (AEAD). Second, the {\it EncryptedSharedPreferences} class is used to automatically encrypt keys and values using a combination of two schemes: first the keys are encrypted using a deterministic algorithm, and then the values are encrypted with {\it AES-256 GCM} in a non-deterministic way.
\subsection{Cryptographic algorithms}
The platform allows you to choose different algorithms for each class. For encryption, it is recommended {\it AES} in {\it CBC} or {\it GCM} mode with 256-bit keys (such as {\it AES/GCM/NoPadding}), for {\it MessageDigest} the {\it SHA-2} family, for Mac the {\it HMAC} of the {\it SHA} family and for signature the {\it SHA-2} family with {\it ECDSA} (such as {\it SHA256withECDSA}).

The execution of different cryptographic operations can be chosen when reading or writing a file, encrypting a message, generating a message digest and generating or verifying a digital signature.
Specifically, there are a multitude of algorithms available that are compatible with Android, such as: {\it DH, DSA, AES, BLOWFISH, ChaCha20, DES, 3DES, EC, GCM, PKCS12PBE, X.509, ECDH, MD5, the SHA family}. In addition, the encryption algorithms {\it AES, AES, AES128, AES256, ARC4, BLOWFISH, ChaCha20, DES, 3DES and RSA} allow you to choose between different modes ({\it CBC, ECB, GCM}), and if desired, you can also choose between different paddings for the chosen algorithm.
\section{Conclusions and future work}
In this work, an application has been designed that collects, through the sensors of a smartwatch, several parameters of some of the most relevant physiological variables that allow early detection if a person is falling asleep while driving. In that case, the application generates an alert that wakes up the driver to avoid possible road accidents. In addition, the implementation has followed the recommended good practices of secure code programming to protect the sensitive data handled by the application.

The possibility of using the BIA sensor is currently being studied in order to use impedance to detect the state of muscle tone, as well as changes in muscle tone, since muscle tone decreases when a person is falling asleep. 
Another possible improvement is the use of the gyroscope sensor to try to detect the wrist dropping motion in the car due to drowsiness.
Also, in order to adapt the alerts to all types of people, two different alarms will be defined: the vibration of the watch and an auditory sensory stimulus with sound.

\section*{Acknowledgments}

This research has been supported by the Spanish Ministry of Science, Innovation and Universities, the State Research Agency and the European Regional Development Fund under project RTI2018-097263-B-I00.
%
% ---- Bibliography ----
%
\bibliographystyle{IEEEtran}

\end{document}